\documentclass[preprint,prb,aps]{revtex4}
\begin{document}
\title{TOWARD A NEW FOUNDATION OF STATISTICAL THERMODYNAMICS.}
\author{J.P. Badiali}
\address{Universit\'{e} P. et M. Curie, Paris France \\
jpbadiali@numericable.com}

\begin{abstract}
We show that in a given number of cases thermodynamic equilibrium states and systems showing irreversible behaviors can be treated on the same footing. This implies two fundamental consequences: the Schr\"{o}dinger equation can not be the corner stone on which statistical thermodynamics is found and stable equilibrium states can be described from a dynamical point of view. We propose to found statistical thermodynamics on a transition function defined in terms of path integral and verifying a time-irreversible equation. The transition function is a weighted sum of paths joining two points in a non-relativistic space-time. Focusing on closed paths we define a probability and characterize the paths by a volume, a kinetic energy and a time. This time of quantum origin is a property of the equilibrium states and it defines a natural unit of time. From the calculations of energy fluctuations on the paths a link with equilibrium thermodynamics is established and all the results previously obtained by the Gibbs ensemble method are re-obtained for large systems and for systems in contact with a thermostat. Replacing the thermostat by a bath without new basic assumption we introduce a friction phenomena. This approach is different from the so called system+reservoir approach since the irreversibility is not due to the bath and a smooth transition from quantal to classical world is established. An extension of our approach is proposed to make a link with systems obeying to a hamiltonian mechanics and exhibiting a time reversible behavior. From this we see that the link between statistical thermodynamics and the pure quantum mechanics described via the Schr\"{o}dinger equation is based on an analysis of time reversibility/irreversibility. The strategy developed in this work consists in using simple models for which results can be derived as exactly as possible and the derivation of a Smoluchovski equation is considered as the signature of the existence of an irreversible process. More general descriptions will not change the physical nature of the results.

\end{abstract}
\maketitle
\vspace{0.5cm}

\section{Introduction}
Thermodynamics is characterized by a very high degree of generality; many branches in physics are independent except that their results must be in accordance with the two fundamental laws of thermodynamics. A part of statistical mechanics is concerned by the calculation of thermodynamic properties starting from a description at a microscopic level, this requires to introduce a hamiltonian model for the system under investigation and to use a particular scheme of calculation including some approximations, in general. This part of statistical mechanics that we can call statistical thermodynamics $(ST)$ is frequently associated with thermodynamics but it cannot be identified with it. For instance, the second law of thermodynamics can be derived from very general arguments without any reference to the existence of a hamiltonian [\cite{lieb}]. If $ST$ adds something very useful to thermodynamics it neither explains thermodynamics nor replaces it. \\
The part of $ST$ firmly founded concerns the description of equilibrium states. In a large class of systems there is a tendency to evolve toward states that are time independent they are are called equilibrium states [\cite{callen}]. A new foundation of $ST$ must reproduce the well known theoretical results concerning these states.
Besides the equilibrium states thermodynamics predicts via the second law that the passage from an equilibrium state $A$ to another one $B$ is only possible if the entropy does not decrease in this transformation. In nature no process is entirely free from friction, heat dissipation, diffusion .... [\cite{planck}], consequently all the processes exhibit an increase of entropy and then the transformation $B \rightarrow A$ is impossible. Reversible processes form only a limiting case of considerable importance for theoretical demonstrations. They are defined as a dense ordered succession of equilibrium reversible states whereas a real process is a temporal succession of equilibrium and non-equilibrium states [\cite{callen}].
Hereafter our main objective is to show that equilibrium states and irreversible processes can be described on the same footing $i.e.$ without introducing new basic assumptions.This leads to a strong consequence; since the Schr\"{o}dinger equation only describes reversible evolutions it cannot be the corner stone from which we can establish a new foundation of $ST$. \\     
From the pioneer work of Boltzmann and the introduction of the H-theorem the question of irreversibility is all the time present in statistical mechanics (see for instance [\cite{zeh}]). The problem initiated by Boltzmann can be summarized as follows: how to go from a microscopic description based on the existence of a hamiltonian leading to time-reversible processes to macroscopic observations that are time-irreversible. Instead of this traditional route to irreversibility we suggest a totally different point of view.  
The hamiltonian $H$ remains the basic quantity defining the system under investigation but any thermodynamic quantity is calculated via a transition function that is a time-dependent transformation of $H$. This function characterizing the paths joining two points in a non-relativistic space-time verifies a time-irreversible differential equation at a microscopic level. This means that the dynamic processes associated with the transition function can be used to re-derive the well known results concerning the equilibrium states. In some sense the problematic of Boltzmann is now inverted: we start from a dynamics that is time-irreversible at a microscopic level and we have to describe stable equilibrium states while the description of non-reversible processes at macroscopic level will be easy. Amongst the numerous sources of irreversibility like friction, diffusion, heat dissipation, electric current .... hereafter we will focus on a friction process generated by a brownian process. This represents a very large class of phenomena and it has been extensively investigated in the literature. \\  
Finally, the last question to investigate is the relation between $ST$ and pure mechanical systems in which there is no friction. For such systems we have a hamiltonian dynamics and the evolution is time-reversible. More generally this leads to investigate the link between $ST$ and quantum mechanics $(QM)$ described at the level of the Schr\"{o}dinger equation. We will see how the time reversibility/irreversibility is crucial in the analysis of this link. \\
The paper is organized as follows. In Section $2$ we select the equilibrium results that a new foundation of statistical mechanics must verify. In Section $3$ we introduce a transition function that gives a counting of paths joining two points in a non-relativistic space-time and we give some properties of this function. In particular we show that it verifies a time-irreversible equation and from it we may introduce two kinds of probabilities; one of them is given via a Smoluchovski equation. In Section $4$ we investigate the dynamical properties associated with closed paths : mean volume of paths, kinetic energy over paths, time to explore the paths. From these results we derive the existence of stable equilibrium states for which thermodynamic properties are identical to those obtained by the traditional Gibbs ensemble method. In the two next Sections we consider a system interacting with a thermostat or a bath. The coupling with a thermostat leads to well known results concerning systems at equilibrium. The coupling with a bath induces friction forces and a time-irreversible behavior. We show that the coupling with these two kinds of environments are essentially due to a quantitative difference between the parameters and no new basic assumption is needed to produce equilibrium states or time-irreversible behaviors. In Section $6$ we show how to extend our approach in order to describe a reversible behavior corresponding to a pure quantum mechanical system. A Schr\"{o}dinger equation will be derived allowing to link $ST$ and $QM$ from a fundamental point of view. The conclusions of this work are given in the last Section.\\
Our approach is restricted to systems in absence of gravity and radiations and theoretical questions as the existence of a thermodynamic limit will be not discussed at this stage. In many places we illustrate our approach by considering simple examples for which analytical results can be derived. The existence of a Smoluchovski equation will be considered as typical from an time-irreversible behavior. A very preliminary version of this work is published in ArXiv [\cite{jpbarxiv}].

\section{Results for equilibrium states}

\subsection{Traditional approach}
The main goal of equilibrium statistical mechanical consists to predict the properties of systems composed of a very huge number of particles; only in the so-called thermodynamic limit a comparison between thermodynamics and $ST$ [\cite{ruelle}] is expected. Since the description of a system at atomic scale requires quantum mechanics it follows that the proper formulation of statistical mechanics must be in quantum mechanical language. \\
It is well known that the expectation of a given quantity is given via the density matrix defined according to [\cite{feyn2}]
\begin{equation}
\rho = \sum_{i} w_{i} \left| i \right\rangle \left\langle  i\right|   
\label{densmat}
\end{equation}
in which the set $ \left|i \right\rangle$ is a complete orthogonal set of vectors and $w_{i}$ defined such that $w_{i} \geq 0$ and $\sum_{i} w_{i} =1$, $w_{i}$ is interpreted as the probability that the system is in a state $i$. Given an operator $A$ the expectation of $A$ is given by 
$\left\langle A \right\rangle = Tr(\rho A) = \sum_{i} w_{i}  \left\langle i \left| A \right| i \right\rangle $  
where $Tr(B)$ means that we have to take the trace of the operator $B$.\\ 
In $ST$ we take for $\left| i \right\rangle$ the eigenfunction of the Hamiltonian $H$ of the system for which the corresponding eigenvalue is $E_{i}$ and we assume that the probability $w_{i}$ to be in a state $i$ is proportional to $exp -\beta E_{i}$ leading to a density matrix in the form
\begin{equation}
\rho= \frac{exp -\beta H}{Tr(exp-\beta H)} = \frac{exp -\beta H}{Q}                                     
\label{canonic}
\end{equation}
where                                                                                      
\begin{equation}
Q = Tr(exp-\beta H)
\label{Q}
\end{equation}
is the partition function and $\beta$ is determined by establishing the connection between $ST$ and thermodynamics.

\subsection{Path integral approach}
Besides this traditional approach Feynman [\cite{feyn2}] observed that the quantity $\rho(u) = \exp \frac{-Hu}{\hbar}$ where $u$ has been redefined to be $\beta \hbar$ verifies the differential equation 
\begin{equation}
\hbar \frac{\partial \rho(u)}{\partial u}= - H \rho(u)
\label{equafif}
\end{equation}
that is formally a Schr\"{o}dinger equation in which we have made the transformation $it = u$. This remark implies that the partition function can be expressed in term of a path integral as it has been done for the wave function. We introduce [\cite{feyn2}]
\begin{equation}
Q_{path}= \int dx^{N}(0) \int \mathcal{D} x^{N}(t) exp(- \frac{1}{\hbar} \int ^{\beta \hbar}_{0} H(s) ds )
\label{partfunc}
\end{equation}
$x^{N}(0)$ represents the set of the positions $x^{i}(0)$ occupied by the $N$ particles at the time $t_{0}$ and $x^{N}(t)$ is a similar quantity but associated with the time $t$, $\mathcal{D} x^{N}(t)$ is the path integral measure. The passage from $Q_{path}$ to $Q$ requires the existence of a multiplicative constant $Cte$ in order to form from $Q_{path}$ a dimensionless quantity, hereafter we consider that all the physics is contained in $Q_{path}$. In Eqs. (\ref{partfunc}) we must take $x^{i}(0) = x^{i}(\beta \hbar)$ and $H(s) = K(s) + U(s)$ where $K(s)$ is given by   
\begin{equation}
K(s) = \sum^{N}_{i=1} \frac{1}{2}m (\frac{dx^{i}(s)}{ds})^{2} = \sum^{N}_{i=1} K^{i}(s)  
\label{K}
\end{equation}
and
\begin{equation}
U(s) =\sum^{N}_{i=1}[v_{1}(x^{i}(s),s) + \sum^{N}_{j=1; j \neq i} v_{2}(x^{i}(s), x^j(s)] =\sum^{N}_{i=1} U^{i}(s)
\label{U}
\end{equation} 
where $v_{1}(x^{i}(s),s)$ is the external potential acting on the particle $i$ located at the point $x^{i}(s)$ at the time $s$ and $v_{2}(x^{i}(s), x^j(s))$ is the pair potential between two particles $i$ and $j$ at time $s$ and $U^{i}(s)$ is the total potential on the particle $i$ located at the point $x^{i}$ at the time $s$. \\ 
Of course from Eqs. (\ref{partfunc}) and Eqs. (\ref{Q}) we will get an identical value for $Q$. However the expressions of $Q$ have to be corrected in order to introduce the particle undiscernability. In [\cite{feyn1}] it has been shown that excepted the case of helium at very low temperature that we exclude here the effect of undiscernability is restricted to the introduction of a factor $\frac{1}{N!}$ in front of $Q$. \\
In the limit $\hbar \to 0$ and at the lowest order in the thermal de Broglie wavelength given by $\Lambda = [\frac{h^{2}}{2 \pi m k_{B}T}]^{\frac{1}{2}}$ where $m$ is the mass of particles and $k_{B}T$ is the order of magnitude of the thermal kinetic energy we have [\cite{hill}] 
\begin{equation}
Q= \frac{1}{N!} \frac{1}{\hbar^{3N}} \int dx^{N} dp^{N} exp - \beta H(x^{N}, p^{N})
\label{qclassic}
\end{equation}
in which we have introduced the set of particle momentum via $p^{N}$. This result is identical to the one given by Eqs. (\ref{partfunc}) if we take $Cte = \frac{1}{\Lambda^{N}}$.\\
Hereafter we adopt the entropy representation of thermodynamics [\cite{lieb}], [\cite{callen}] for which the basic relation is given by the entropy $S$ as a function of the internal energy 
$U$, the volume $V$ and the number of particles $N$ for a mono component system . Using the relation $F = - k_{B} \ln Q$ between the free energy $F$ and $Q$ and the thermodynamic relation $F = U - TS$ we obtain in terms of path integral
\begin{equation}
S = k_{B} \ln \frac{1}{N!}\int dx^{N}(0) \int \mathcal{D} x^{N}(t) exp(- \frac{1}{h} \int ^{\beta \hbar}_{0} [H(s) - U] ds ) 
\label{spath}
\end{equation}
showing that the entropy is determined by the fluctuations of the internal energy along the paths these fluctuations being such as $\int ^{\tau}_{0}(H(s)-U) ds \approx h$.\\
In what follows starting from a dynamic point of view our task will be to re-derive Eqs. (\ref{qclassic}) and Eqs. (\ref{spath}). 

\subsection{Comments}
The previous well known results require nevertheless two comments. Frequently it is claimed that we can pass from quantum mechanics $QM$ to $ST$ by introducing the imaginary time $u = it$. This is just a mathematical trick maintaining open the following question: why in quantum mechanics to get a physical quantity we have to associate the wave function with its complex conjugate while in $ST$ it is enough to introduce the imaginary time only into the wave function in order to get a physical quantity. Hereafter we will show that the passage from $ST$ to quantum mechanics requires an analysis in term of time reversibility/irreversibility.\\
The second comment has been introduced by Feynman [\cite{feyn1}] who compared the traditional expression of the partition function Eqs. (\ref{Q}) to its path integral form given by   
Eqs. (\ref{partfunc}). He noticed that the traditional approach requires to use the canonical form of $w_{i}$ and to solve the Sch\"{o}dinger equation this means that all the details of the quantum mechanics are needed in order to calculate $Q$. Feynman [\cite{feyn1}] (see '' Remarks on methods of derivation'' p. 295) suggested that it must be possible to derive the path integral expression of $Q$ without solving the Schr\"{o}dinger equation but directly starting from the time-dependent motion. Hereafter we will follow this route. Note that a similar observation has been done in the case of black holes where it has been shown that the black holes thermodynamics does not require the precise form of the Einstein equation [\cite{wald}]. There is a more basic element showing that the Schr\"{o}dinger equation cannot be the corner stone on which we can found $ST$. Thermodynamics also contains via the second law the description of non-reversible processes and if we want to describe equilibrium states and non-equilibrium situations on the same footing we can not start from the Schr\"{o}dinger equation which is limited to the description of reversible processes. 

\section{The transition function and its properties} 
Hereafter we characterize a quantum mono-component system enclosed in a box of volume $V$ by a transition function that will be our basic tool to describe both equilibrium states and irreversible behaviors.

\subsection{Introduction of the transition function}
The hamiltonian $H$ represents our basic knowledge on a system. In classical mechanics from $H$ we can calculate the trajectory of a particle for a given time interval and fixed initial and final positions. For a isolated system with $H$ is associated a constant value for the system energy. In quantum mechanics there is no definite trajectory for a particle but we can calculate the sum of all the paths joining the points $x^{N}(0)$ and $x^{N}$ when the time is running from $t_{0}$ to $t$. Of course, not all the paths have the same importance for a physical system and for the physics we consider therefore we have to introduce a weighted sum over the paths. In quantum physics the natural scale is given by the Planck constant accordingly we decide that the relevant paths are those for which the action does not differ very much from $\hbar$. The action we consider is the euclidean action that eliminates for a given time interval the too high values for the energy. A quantity verifying all these properties is the path integral defined for $t \geq t_{0}$ 
\begin{equation}
\phi(x^{N}(0), t_{0}; x^{N}, t) = \frac{1}{N!} \int \mathcal{D} x^{N}(t) exp(- \frac{1}{\hbar} \int ^{t}_{t_{0}} H(s) ds )
\label{tool}
\end{equation} 
$x^{N}(0)$ represents the set of the positions $x^{i}(0)$ occupied by the $N$ particles at the time $t_{0}$ and $x^{N}$ is a similar quantity but associated with the time $t$, $\mathcal{D} x^{N}(t)$ is the path integral measure and $H(s)$ is the hamiltonian associated with the system. The time $t$ appearing in Eqs. (\ref{tool}) is the usual time $i.e.$ a real quantity and 
$\phi(x^{N}(0), t_{0}; x^{N}, t)$ is a positive real valued function. \\
To calculate the path integral in Eqs. (\ref{tool}) we introduce a time discretization, the time step $\delta t$ is given by $\delta t = \frac{t - t_{0}}{n}$ where $n$ is very large and we denote by $\delta x$ the difference of position corresponding to $\delta t$. The path integral measure is given by 
\begin{equation}
\mathcal{D} x^{N}(t) = \frac{1}{C} \prod^{N}_{i =1} \prod^{n-1}_{j=1} dx^{i}_{j}
\label{measure}
\end{equation}
in which $C$ is a normalization constant and $dx^{i}_{j}$ represents the position of the particle $i$ at the time $t_{0} + j \delta t$, we have $x^{i}_{0} = x^{i}(0)$ and 
$x^{i}_{n-1} = x^{i}(t -\delta t)$. \\
For a particle $i$ having a hamiltonian $H^{i}(s) = K^{i}(s) + U^{i}(s)$ defined via (\ref{K}) and (\ref{U}) we have in the limit $\delta t \to 0$ 
\begin{equation}
limit _{\delta t \to 0}\frac{1}{\hbar}\int ^{t+ \delta t}_{t} H^{i}(s) ds \approx \frac{1}{\hbar} [\frac{1}{2}m \frac{(\delta x^{i})^{2}}{\delta t} + \delta t U^{i}(t))] 
\label{actiondt} 
\end{equation}
where we have assumed that the potential does not vary during $\delta t$. If we take $(\delta x^{i})^{2} \approx \delta t^{1 + \alpha}$ with $\alpha > 0$ the contribution to the path is vanishingly small. In opposite for negative values of $\alpha$ Eqs. (\ref{actiondt}) behaves like $\frac{1}{\delta t^{\alpha}}$ giving a vanishing contribution to the path integral Eqs. (\ref{tool}). Thus  the relevant paths are those for which $\frac{1}{2} m\frac{(\delta x^{i})^{2}}{\delta t} \approx \hbar$ leading to 
\begin{equation}
[\frac{1}{2}m (\frac{\delta x^{i}}{\delta t})^{2} + U^{i}(t)] \delta t = \delta E^{i} \delta t \approx \hbar
\label{dedt}
\end{equation}
Thus during $\delta t$ we accept the fluctuations of energy verifying Eqs. (\ref{dedt}) $i.e.$ a kind of Heisenberg uncertainty relation. In the limit $\delta t \to 0$ the main paths are those for which we have in one dimension  
\begin{equation}
(\delta x^{i})^{2} \approx 2 \frac{\hbar}{m} \delta t
\label{dif}
\end{equation}
mimicking a diffusion process. \\
There is no doubt that $\phi(x^{N}(0), t_{0}; x^{N}, t)$ characterizes a system but it is not the only one possibility a totally different quantity or another definition of $\phi(x^{N}(0), t_{0}; x^{N}, t)$ might be considered. Nevertheless $\phi(x^{N}(0), t_{0}; x^{N}, t)$ given by Eqs. (\ref{tool}) represents a simple choice and below we will examine all the consequences of it. \\
In summary, instead of the hamiltonian which is the traditionnal basis of $ST$ we propose to consider that the basis should be a time dependent transformation of it via the existence of a transition function. 

\subsection{Basic properties of the transition function}   
Associated with the path integral formalism there it exists a large number of results explicitly given in [\cite{feyn1}], [\cite{schulman}], [\cite{kleinert}] for instance . Here we do not recall them but focus on those useful hereafter.\\ 
From the function $\phi(x^{N}(0), t_{0}; x^{N}, t)$  we may define a semi-group since 
\begin{equation}
\phi(x^{N}(0), t_{0}; x^{N}, t) = \int \phi(x^{N}(0), t_{0}; x^{N}(1), t_{1})\phi(x^{N}(1), t_{1}; x^{N}, t) dx^{N}(1)
\label{semigroup}
\end{equation}
whatever $t_{1}$ provided $t_{0} \leq t_{1} \leq t$. In the limit $t \to t_{0}$ we have
\begin{equation}
limit_{t \to t_{0}}\phi(x^{N}(0), t_{0}; x^{N}, t) = \prod^{N}_{i} \delta (x^{i} - x^{i}(0))
\label{dirac}
\end{equation}
From $\phi(x^{N}(0), t_{0}; x^{N}, t)$ we define 
\begin{equation}
\phi(x^{N},t)= \int \phi_{0}(x^{N}(0))\phi(x^{N}(0), t_{0}; x^{N}, t) dx^{N}(0)
\label{deffi}
\end{equation}
in which $\phi_{0}(x^{N}(0))$ is a given function and due to Eqs. (\ref{dirac}) we have $\phi(x^{N},t = t_{0}) = \phi_{0}(x^{N}(0))$. The equations (\ref{semigroup}), (\ref{dirac}) and (\ref{deffi}) show that $\phi(x^{N}(0), t_{0}; x^{N}, t)$ plays the role of a transition function in the space-time.\\
For system in absence of interaction potential we have in one dimension
\begin{equation}
\phi_{ideal}(x^{N}(0), t_{0}; x^{N}, t) = \prod^{N}_{i=1}\phi^{i}_{ideal}(x^{i}_{0}, t_{0}; x^{i}, t) = \prod^{N}_{i=1} (\frac{m}{2 \pi \hbar (t-t_{0})})^{\frac{1}{2}} \exp [-\frac{m}{2 \hbar (t - t_{0})}(x^{i} - x^{i}_{0})^{2}]
\label{fiideal}
\end{equation}

\subsection{Time evolution of the transition function}
In order to investigate the time evolution of the transition function we follow a route similar to the one used by Feynman [\cite{feyn1}] in order to relate the path integral formalism of quantum mechanics to the Schr\"{o}dinger equation. For an infinitely small increasing of time $\epsilon$ we have using Eqs. (\ref{semigroup})
\begin{equation}
\phi(x^{N}(0), t_{0}; x^{N}, t+ \epsilon) = \int \phi(x^{N}(0), t_{0}; x^{N}(1), t)\phi(x^{N}(1), t; x^{N}, t + \epsilon) dx^{N}(1)
\label{semieps}
\end{equation}
In the part $\phi(x^{N}(1), t; x^{N}, t + \epsilon)$ of this expression the dominating term is the kinetic energy proportional to $\frac{1}{\epsilon}$ and the potential energy can be expanded linearly in term of $\epsilon$, due to this we have to deal with gaussian quadratures. In the part $\phi(x^{N}(0), t_{0}; x^{N}(1), t)$ we may expand $x^{N}(1)$ around $x^{N}$. Many terms can be eliminated with the gaussian quadrature and finally selecting all the terms of order $\epsilon$ we are left with (see [\cite{jpb5}] for a detailed demonstration) 
\begin{equation}
\frac{\partial\phi(x^{N}(0), t_{0}; x^{N}, t)}{\partial t} = \frac{\hbar}{2m} \sum^{N}_{i=1} \frac{\partial^{2} \phi(x^{N}(0), t_{0}; x^{N}, t)}{\partial (x^{i})^{2}} 
- \frac{1}{\hbar} \sum^{N}_{i=1} U^{i}(t)\phi(x^{N}(0), t_{0}; x^{N}, t) 
\label{dfidt1}
\end{equation}  
 Introducing the function $\phi(x^{N},t)$ defined by Eqs. (\ref{deffi}) we obtain
\begin{equation}
\frac{\partial\phi(x^{N},t)}{\partial t} = \frac{h}{2m} \sum^{N}_{i=1} \frac{\partial^{2} \phi(x^{N},t)}{\partial (x^{i})^{2}} 
- \frac{1}{\hbar} \sum^{N}_{i=1} U^{i}(t)\phi(x^{N},t) 
\label{dfidt2}
\end{equation}
assuming that the transition function is regular in order to be able of permuting derivative and quadrature. \\
In a previous work [\cite{jpb2}] the equation (\ref{dfidt2}) has been established from a semi-empirical point of view inspired by the chess-model [\cite{feyn2}],[\cite{ord1}]; the space time was considered as initially discrete but a continuous limit was sufficient for the investigated cases. In this route we started from (\ref{dfidt2}) and the path integral form of the transition function Eqs. (\ref{tool}) was obtained via the Feynman-Kac formula. Here a more fundamental point of view is adopted: the space time is continuous but a discretization appears from the process of quantification associated with the path integral.  

\subsection{Probabilities associated with the transition function}
The equation (\ref{dfidt2}) is a diffusion type equation due the existence of a paths selection verifying Eqs. (\ref{dif}). However the potential term is equivalent to introduce a creation/annihilation 
effect and $\phi(x^{N}, t)$ can not be not normalized and it can not define a probability. Nevertheless from the transition function it is possible to create two probabilities.

\subsubsection{Path probability}
Let consider 
\begin{equation}
p(x^{N}(0), t_{0}; x^{N}, t) = \frac{exp- \frac{1}{\hbar} \int^{t}_{t_{0}} H(s) ds}{\int \mathcal{D} x^{N}(t)exp- \frac{1}{\hbar} \int^{t}_{t_{0}} H(s) ds}
\label{denspathproba}
\end{equation}
this quantity is positive, smaller than $1$ and normalized since $\int \mathcal{D} x^{N}(t)p(x^{N}(0), t_{0}; x^{N}, t) =1$. We  consider $p(x^{N}(0), t_{0}; x^{N}, t)\mathcal{D} x^{N}(t) $
as the probability to have paths included in the volume $\prod^{N}_{i =1} \prod^{n-1}_{j=1} dx^{i}_{j}$ around the set of points $[x^{k}_l)]$. For a given function $A([x^{k}_l])$ we can define an average over paths according to
\begin{equation}
<A>_{path} = \int  \prod^{N}_{i =1} \prod^{n-1}_{j=1} dx^{i}_{j} A([x^{k}_l])p(x^{N}(0), t_{0}; x^{N}, t)
\label{avepath}        
\end{equation}
The density of probability Eqs. (\ref{denspathproba}) has already been used to calculated the properties of closed paths [\cite{jpb0}]. The average over paths is a special procedure that we can not reduce to a time average or to a traditional canonical average, in general. To illustrate the calculation of an average let consider a quantity $A$ defined in $t_{l}$ and $t_{l+1}$ such as $t_{0} \leq t_{l} < t_{l+1} \leq t$. For a given path in $t_{l}$ and $t_{l+1}$ the positions are $x_{l}$ and $x_{+1}$ respectively and from them we can form 
$A(x_{l},x_{l+1})$. Finally we perform a sum of all the $A(x_{l},x_{l+1})$ on all the paths weighted by $p(x^{N}(0), t_{0}; x^{N}, t)$; the result depends on $x^{N}(0)$, $x^{N}$ , $t_{0}$ 
and $t$. In what follows Eqs. (\ref{denspathproba}) and Eqs. (\ref{avepath}) are the ingredients on which our approach of $ST$ is based.  

\subsubsection{Smoluchovski equation}
From $\phi(x^{N}, t)$ we may define another kind of probability. To have a simple analytical result we consider the case of non-interacting particles located in a time-independent external potential. In that case the transition function is factorized in a product of functions $\phi(x,t)$ associated with one particle in the external potential $U(x)$. For $\phi(x,t)$ we follow the mathematical transformation introduced in [\cite{kampen}] and we consider the new quantity $P(x,t) = Z(x) \phi(x,t)$. By using Eqs. (\ref{dfidt2}) and straightforward manipulations we get the following results
\begin{equation}
\frac{\partial P(x,t)}{\partial t} = \frac{\hbar}{ 2m}\frac{\partial ^{2}P(x,t)}{\partial x^{2}} + \frac{\hbar}{m} \frac{\partial}{\partial x}[\frac{\partial V(x)}{\partial x} P(x,t)]
\label{smol}
\end{equation}
in which the potential $V(x)$ is defined by $Z(x) = \exp- V(x)$ and $Z(x)$ is the solution of
\begin{equation}
- \frac{\hbar}{2m} \frac{\partial^{2}Z(x)}{\partial x^{2}} + U(x) Z(x) = 0
\label{defZ}
\end{equation}
$P(x,t)$ verifying the Smoluchovski equation (\ref{smol}) defines a probability for which we have an equilibrium probability $P_{eq}(x)$ corresponding to $\frac{\partial P(x,t)}{\partial t} = 0$ and given by $P(x) = C \exp - V(x) $ in which $C$ is a normalization constant. Thus from $\phi(x^{N}(0), t_{0}; x^{N}, t)$ we may deduce a density of probability verifying in simple situations a Smoluchovski equation. We may expect that in more complicated situations it must be also possible to define a probability. If the external potential is time dependent a similar result can be obtained replacing $Z(x)$ by $Z(x,t)$ and an equation similar to Eqs. (\ref{defZ}) can be derived. If we introduce an interaction between particles a general result can be obtained for instance by a perturbation method leading to a modification of Eqs. (\ref{smol}). This shows that $\phi(x^{N}(0), t_{0}; x^{N}, t)$ is sufficient to define a quantity i)having a physical meaning and ii) presenting a time-irreversible behavior. \\
In a recent paper [\cite{jpb1}] the results presented in this section have been used to describe an irreversible chemical reaction in vacuum and it has been shown that the stochastic approach of the chemical reaction proposed by Kramers [\cite{kramers}] can be justified from first principles. $P(x,t)$ is then interpreted as the density of probability for particle to be at the position $x$ at the time $t$.  

\section{Dynamical description of the thermodynamic equilibrium}
For a system $S$ in thermodynamic equilibrium all its properties are time-independent, by definition. In principle, any experimental measurement on $S$ is the result of a time averaging process but in the Gibbs ensemble method this averaging is replaced by an average on an ensemble of systems having a structure similar to the one of $S$ [\cite{hill}] and there is no time in the calculation of equilibrium values. If we start from the dynamical point of view associated with the transition function the system is all the time fluctuating around its equilibrium values. To extract a thermodynamic data from a fluctuating system needs to consider a given time in which all the fluctuations are averaged. This time depends both on the apparatus we use and on the system specificities moreover it must be finite since we want to be able to reproduce several measurements in a finite period of time. \\
In the calculation of $\phi(x^{N}(0), t_{0}; x^{N}, t)$ the role of time depends on the class of paths we investigate. Closed paths are defined by the space point from which they start and return but, in addition, to perform a calculation we have to specify the time interval during which the paths are explored. If we may take the same time interval whatever the initial position of the paths we introduce an internal parameter that we have to determine in order to have a self contained approach. If we focus on a class of open paths joining the point $x_{0}$ to $x$ we have a description that contains time-dependent correlation functions. These functions give information beyond what we know from thermodynamics. For such functions we do not see how to eliminate $t$ excepted by an integration over all the value of $t$ running from $0$ to $\infty$. This route requires at least in principle an infinite time to measure a quantity and due to this we decide to stay with closed paths. This choice is also inspired by what is done in traditional quantum statistical physics where the properties of thermal equilibrium are related to the trace of the density matrix, more sophisticated descriptions used the non-diagonal terms of the density matrix. Below several quantities are introduced to characterize the paths.

\subsection{Mean volume associated with closed paths}
In equilibrium situation the transition function associated with two times $t$ and $t_{0}$ must only depend on $(t-t_{0}) =\tau$ and we can take $t_{0} = 0$. Explicit calculations are performed on a one-dimensional system, a generalization to three dimensions is straightforward.\\   
Let consider first a system without interaction then $\phi(x^{N}(0), t_{0}; x^{N}, t) = \prod^{N}_{i=1}\phi^{i}_{ideal}(x^{i}_{0}, 0; x^{i}, \tau)$ in which all the functions $\phi_{ideal}^{i}(x^{i}_{0}, 0; x^{i}, \tau)$ are similar and given by Eqs. (\ref{fiideal}). In this case we are free to explore all the paths during the same time interval $\tau$.  For a given point $x^{i}(0)$ the quadratic deviation around this point is calculated using Eqs. (\ref{avepath}) leading to     
\begin{equation}
<[x^{i}(t') - x^{i}(0)]^{2}>_{path} =  \int \prod ^{n-1}_{j =1}dx^{i}_{j}[x^{i}(t') - x^{i}(0)]^{2} p_{ideal}(x^{i}(0), 0; x^{i}, \tau)
\label{x2}
\end{equation}
where $0 < t' < \tau$ and we have used the fact that the density of probability Eqs. (\ref{denspathproba}) can be factorized.  The calculation of Eqs. (\ref{x2}) it reduces to gaussian quadratures and we obtain 
\begin{equation}
<[x^{i}(t') - x^{i}(0)]^{2}>_{path} =\frac{\hbar}{m} \frac{t'(\tau - t')}{\tau}
\label{x2result}
\end{equation}
 that exhibits a maximum for $t' = \frac{\tau}{2}$.  leading to 
\begin{equation}
max <[x^{i}(t') - x^{i}(0)]^{2}>_{path} = \frac{\hbar \tau}{4m} = \lambda^{2}
\label{max}
\end{equation} 
Thus a free particle $i$ can move in a sphere of radius $\lambda = \sqrt{\frac{\hbar \tau}{4m}}$. In presence of a potential $U^{i}(t')$ we assume that the variation of this potential is negligible in this sphere and $\lambda$ is not changed. We will see that this point corresponds to a standard assumption [\cite{hill}] in order to derive Eqs. (\ref{qclassic}) that is our goal.

\subsection{Kinetic energy on the paths}
In Eqs. (\ref{x2result}) we can choose $t' = \delta t$ and consider the kinetic energy $<e_{K}>_{path}$ defined as
\begin{equation}
<e_{K}>_{path} = \frac{1}{2} m <\frac{(x^{i}(\delta t) - x^{i}(0))^{2}}{\delta t^{2}}>_{path} = \frac{\hbar}{2}[\frac{1}{\delta t} - \frac{1}{\tau}] = \frac{\hbar}{2 \delta t} - <E_{K}>_{path}
\label{ekpath}
\end{equation}
In the limit $\delta t \to 0$ this kinetic energy is positive as expected but it grows like $\frac{1}{\delta t}$. This behavior is associated with the existence of a quantum of elementary action; we have $<e_{K}>_{path}  \delta t = \frac{\hbar}{2}$. The finite contribution to $<e_{K}>_{path}$ is negative and we have $<E_{K}>_{path} = \frac{\hbar}{2 \tau}$. The results show that an average over paths - that is the only tool we have at our disposal - may lead to non-traditional results as already show in [\cite{jpb0}].    

\subsection{Characteristic time for equilibrium states} 
Let analyze the fluctuations of energy over the closed paths via a quantity $S(\tau)$ similar to Eqs. (\ref{spath}) but depending on a time interval $\tau$ that is an unknown parameter for the moment. We define 
\begin{eqnarray}
&&S(\tau)=k_{B} \ln \frac{1}{N!}\int dx^{N}(0)\int \mathcal{D} x^{N}(t) exp(- \frac{1}{\hbar}\int ^{\tau}_{0}(H(s)-U) ds) \nonumber\\ 
&& = \frac{k_{B}\tau U}{\hbar}+ k_{B} \ln\int dx^{N}(0)\phi(x^{N}(0),0; x^{N}(0),\tau) 
\label{S}
\end{eqnarray} 
$U$ is the internal energy needed to create the system. To determine $\tau$ we consider the derivative $[\frac{\partial S(\tau)}{\partial U}]_{N,V}$ given by
\begin{equation}
[\frac{\partial S(\tau)}{\partial U}]_{N,V} = k_{B} [\frac{\tau}{\hbar} + [\frac{\partial \tau}{\partial U}]_{N,V}[\frac{U}{\hbar} + [\frac{\partial \ln \int dx^{N}(0)\phi(x^{N}(0),0; x^{N}(0),\tau)}{\partial\tau}]_{N,V}]]  
\label{deriv1}
\end{equation}
where we have taken into account that the transition function depend on $U$ only via $\tau$. In order to use the time evolution equation (\ref{dfidt1}) in the calculation of the partial derivative of $\phi(x^{N}(0),0; x^{N}(0),\tau)$ we rewrite this function using the semi-group property Eqs. (\ref{semigroup})
\begin{equation}
\phi(x^{N}(0),0; x^{N}(0),\tau) = \int dx'^{N} \phi(x^{N}(0),0; x'^{N},\delta t) \phi(x'^{N},\delta t; x^{N}(0),\tau - \delta t)
\label{dfidto1}
\end{equation} 
and the derivative is given by 
\begin{equation}
\frac{\partial \phi(x^{N}(0),0; x^{N}(0),\tau)}{\partial \tau} = \int dx'^{N} \phi(x^{N}(0),0; x'^{N},\delta t) \frac{\partial \phi(x'^{N},\delta t; x^{N}(0),\tau - \delta t)}{\partial \tau}
\label{dfidto2}
\end{equation}
or using Eqs. (\ref{dfidt1}) the $r.h.s.$ of Eqs. (\ref{dfidto2}) becomes  
\begin{equation}
\int dx'^{N} \phi(x^{N}(0),0; x'^{N},\delta t) [\frac{\hbar}{2m} \sum^{N}_{i=1} \frac{\partial^{2} \phi(x'^{N}, \delta t; x^{N}, \tau - \delta t)}{\partial (x^{i})^{2}} 
- \frac{1}{\hbar} \sum^{N}_{i=1} U^{i}(\tau - \delta t)\phi(x'^{N},\delta t; x^{N}(0),\tau - \delta t))]
\label{dfidto3}
\end{equation}
The part associated with the interaction potential gives the following contribution to Eqs. (\ref{deriv1})  
\begin{equation}
<Up(\tau)>_{path} = \lim_{\delta t \to 0} \frac{\int dx^{N}(0) \int dx'^{N} \phi (x^{N}(0),0; x'^{N}, \delta t) \sum^{N}_{i=1} U^{i}(\tau - \delta t) \phi(x'^{N}, \delta t; x^{N}, \tau - \delta t)}
{\int dx^{N}(0) \phi(x^{N},0; x^{N}, \tau)}
\label{up1}
\end{equation}
In the limit $\delta t \to 0$ we have $U^{i}(\tau - \delta t) \approx U^{i}(\tau) \approx U^{i}(0)$ the last approximation results from our hypothesis that during $\tau$ the particle explores a volume of radius $\lambda$ in which there is no significant variation of potential. Finally we can write
 \begin{equation}
<Up(\tau)>_{path} = \frac{\int dx^{N}(0) \sum^{N}_{i=1}U^{i}(0) exp -(\frac{\tau}{\hbar}\sum^{N}_{j=1}U^{j}(0))}{ \int dx^{N}(0)exp -(\frac{\tau}{\hbar}\sum^{N}_{j=1}U^{j}(0))} 
\label{up2}
\end{equation}
that we interpret as the mean value of the potential energy calculated over the closed paths. \\
The calculation of the first term in Eqs. (\ref{dfidto3}) requires to calculate the second derivative of $\phi(x'^{N}, \delta t; x^{N}, \tau - \delta t)$. This can be readily performed if we use the discrete form of the transition function and then if we write $x'^{i} = x^{i} + \delta x^{i}$. Straightforward calculations lead to   
\begin{equation}
 \lim_{\delta t \to 0} \frac{\int dx^{N}(0) \int dx'^{N} \phi (x^{N}(0),0; \prod^{N}_{i=1}(x^{i} + \delta x^{i}), \delta t) \sum^{N}_{i=1}(\frac{m}{2} (\frac {\delta x^{i}}{\delta t})^{2} -\frac{h}{2 \delta t})
\phi(\prod^{N}_{i=1}(x^{i}+ \delta x^{i}), \delta t; x^{N},\tau - \delta t)}{\int dx^{N}(0) \phi(x^{N},0; x^{N}, \tau)}
\label{uk}
\end{equation} 
where the potential contribution leads to a term of order $\delta t$ that we can neglect. Thus Eqs. (\ref{uk}) is nothing else than of $<(\frac{m}{2} (\frac {\delta x^{i}}{\delta t})^{2} -\frac{h}{2 \delta t})>_{path}$. In this average over paths the first term is $<e_{K}>_{path}$ given in Eqs. (\ref{ekpath}) and we see that its diverging contribution is canceled by the second term. Finally we can write using Eqs. (\ref{ekpath})
\begin{equation}
<(\frac{m}{2} (\frac {\delta x^{i}}{\delta t})^{2} -\frac{h}{2 \delta t})>_{path} = -<Uk(\tau)>_{path} = - \frac{N \hbar}{2 \tau}
\label{ekfinite}
\end{equation} 
We can rewrite Eqs. (\ref{dfidto2}) according to
\begin{equation}
[\frac{\partial S(\tau)}{\partial U}]_{N,V} = k_{B}[\frac{\tau}{h} + \frac{1}{h}[\frac{\partial \tau}{\partial U}]_{N,V}[U - (<Uk(\tau)>_{path} + <Up(\tau)>_{path})]]  
\label{derivfinal}
\end{equation}
The sum $<U(\tau)>_{path} = <Uk>_{path} +<Up>_{path}$ is a finite quantity associated with the average of energy calculated over the paths. A natural equilibrium condition consists to choose $\tau$ in order to have
\begin{equation}
U - <U(\tau)>_{path} = 0
\label{equi}
\end{equation} 
This condition gives a relation between $U$ and $\tau$. As noted in Section $3.D.1$ an average over paths is not a time average. Here it means that the energy -essentially the kinetics energy- must be averaged on paths that we must explore during $\tau$. Thus from $\tau$ we introduce a unit of time characterizing an equilibrium state. If a measurement on a system is performed with an ideal apparatus but on a time interval time $t < \tau$ the result will be unpredictable as a result of quantum fluctuations. Now Eqs. (\ref{deriv1}) is reduced to 
\begin{equation}
[\frac{\partial S(\tau)}{\partial U}]_{N,V}= \frac{k_{B}\tau}{\hbar}= \frac{1}{T_{path}}
\label{deriv2}
\end{equation}   
in which we have introduced a temperature over the paths $T_{path}$ and since $\tau >0$ we can conclude that $T_{path} > 0$. 

\subsection{Relation with thermodynamics}
In the previous subsections the dynamics associated with the closed paths has been characterized by several quantities: the mean volume explored by the paths, the elementary kinetic energy and a characteristic time. Starting from $S(\tau)$ characterizing the energy fluctuations over the paths we have established the relation Eqs. (\ref{deriv2}). As usually in standard $ST$ we can try to relate the calculated results with those of thermodynamics by identifying two similar quantities from which we have to recover relations  existing simultaneously in thermodynamics and $ST$. Let assume that the path temperature $T_{path}$ is identical to the usual thermodynamic temperature. A lot of result can be deduced from this assumption. First from Eqs. (\ref{deriv2}) we deduce 
\begin{equation}
\tau = \frac{\hbar}{k_{B} T} = \beta \hbar
\label{tau}
\end{equation}
Now $S(\tau)$ defined in Eqs. (\ref{S}) is identical to the $S$ defined in Eqs. (\ref{spath}) that is identical to the expression of the entropy derived from the Gibbs ensemble method. We know that the entropy defined as a function of $U$ and $N$ is the fundamental relation in thermodynamics [\cite{callen}] and thus from our approach we will recover all the results obtained via the Gibbs ensemble method. The relation Eqs. (\ref{deriv2}) is now the usual definition of temperature via the entropy. The quantity $<Up(\tau)>_{path}$ given in Eqs. (\ref{up2}) is now identical to the mean value of the potential calculated in the canonical ensemble. For the kinetic energy we have $<Uk(\tau)>_{path} =  \frac{N \hbar}{2 \tau} = N \frac{k_{B}T}{2}$ which is the usual value of the thermal kinetic energy in one dimension. The condition of equilibrium Eqs. (\ref{equi}) is nothing else than the expression of the internal energy in terms of microscopic properties. 

\subsection{Comments}
The time $\tau$ appears frequently in the literature, for instance, in [\cite{gibbons}] it has been used without justification for calculating the black hole radiation. It is interesting to mention that the relation between equilibrium states and dynamics has been already investigated in the literature. It has been shown that an equilibrium state represented by a density matrix induces the existence of a dynamics characterized by the usual evolution operator provided that the time is rescaled by $\tau$ that is considered as the natural unit of time [\cite{connes}] but the origin of this scaling is not elucidated. Above we have seen why $\tau$ can be considered as the unit of time. The results developed in [\cite{connes}] for finding a thermodynamics origin of time in presence of gravity used a definition of the thermal equilibrium similar to (\ref{equi}).\\ 
The uncertainty relation Eqs. (\ref{dedt}) shows that for $\delta t < \tau$ the energy fluctuations of quantum origin are larger than the thermal ones represented by $k_{B}T$ and we can conclude that for this time interval there is no thermodynamics. The non-existence of thermodynamics for very short period of time has been already underlined by Landau [\cite{landau}] here we give a quantitative version to the Landau remark. This point is also very satisfying for us because it allows us to conciliate the description of an equilibrium state with the existence of a time-irreversible differential equation. On each closed path there is no heat dissipation since heat has no meaning for such short time intervals and we may conclude that the results obtained in this Section correspond to stable thermodynamic equilibrium states. \\
If a measurement is done on a time interval shorter than $\tau$ the result will be unpredictable. $\tau$ corresponds to the unit of time in the measurement of a thermodynamic quantity as already suggested in [\cite{connes}] and [\cite{jpb4}]. The value of $\tau$ is $0.76 10^{-11}\frac{1}{T}sec$ and for an experimental measurement performed in $1\mu sec$ for instance we have $\approx 10^{5} T$ units of time and there is no doubt that the experiment will give the exact theoretical average. More drastically $\tau$ has a quantum origin since related to $\hbar$ and in a classical limit in comparison with time required for experimental measurements we can take $\tau \to 0$. However we can not take this limit in any circumstance because as illustrated below 
Eqs.(\ref{classic}) we may have to consider the ratio $\frac{\tau}{\hbar}$. \\   
We have assumed that the variations of $U^{i}$ on a distance $\lambda$ from the center of a particle $i$ can be neglected this also corresponds to the traditional assumption [\cite{hill}] since $\lambda \approx \Lambda$. In particular this assumption was needed to describe the particle undiscernability by just a factor $\frac{1}{N!}$ (\cite{feyn1}). In nanometer we have $\lambda = 0.5 (\frac{m_{p}}{m}\frac{1}{T})^{\frac{1}{2}}$ where $m_{p}$ is the proton mass. This distance has to be compared with the core size $\sigma$ of a particle $i$. Excepted the case of very light particles at very low temperature - that we have eliminated from the very beginning - we wee that $\lambda \leq \sigma $. In the core region $U^{i}$ is much smaller than the core potential and the approximation is verified in a large number of cases. In contrast on $\lambda$ we have to treat carefully the kinetics energy which is large and associated with the fractal character of paths
[\cite{jpb0}].\\
Another consequence of the dynamical approach of equilibrium is to introduce a relation between action and entropy, such relation has been verified in the case of Schwatzchild black holes [\cite{jpb3}].

\subsection{The classical result}
During the time interval $\tau$ the fluctuations in the positions of a free particle are restricted to a volume of radius $\lambda \approx \Lambda$. Since we have assumed that the variations of $U^{i}$ are negligible on a sphere of radius $\lambda$ we have 
\begin{equation}
\frac{1}{\hbar} \int ^{\tau}_{0} U^{i}(s)ds \approx \beta U^{i}(0)
\label{classic}
\end{equation} 
and we have  essentially to focus on the kinetic energy in a closed paths. To do that we may divide the time interval $\tau$ in two equal time intervals and we change $x^{i}(1)$ in $v^{i}$ according to $x^{i}(1) = x^{i}(0) + \frac{\tau}{2}v^{i}$ we get 
\begin{equation}
Q_{path} = \int dx^{N}(0)\phi(x^{N}(0),0; x^{N}(0),\tau) = \frac{\Lambda^{N}}{N! h^{3N}} \int dx^{N}(0) exp(-\beta \sum^{N}_{i=1}U^{i}(0))\int \prod^{N}_{i=1}dp^{i} exp -\beta \frac{(p^{i})^{2}}{2 m}  
\label{classical}
\end{equation} 
which is identical to Eqs. (\ref{qclassic}) if we take $Q = Cte Q_{path}$ with $Cte = \frac{1}{\Lambda^{N}}$.
   
\section{System in equilibrium with a thermostat}
In the previous Section we considered a very large system for which the thermodynamic properties are well defined. Now we analyze the properties of a system $S$ enclosed in a box of volume $V$ outside of which there is a very large system $\mathcal T$ having a volume $\mathcal V >> V$. Our first task work consists to write Eqs. (\ref{tool}) for two systems in contact. \\
The hamiltonian $H(s)$ becomes $H(s) = H_{S}(s) + H_{\mathcal T}(s) + I(s)$ in which $H_{S}(s)$ and $H_{\mathcal T}(s)$ are associated with the isolated systems and $I(s)$ to their interaction. The position of a particle $\alpha$ of $\mathcal T$ will be referred by $y^{\alpha}$, their total number is $M$ and their mass is $m_{\alpha}$. The interaction energy $I(s)$ can be written $I(s) = \sum^{N}_{i=1} \sum^{M}_{\alpha = 1} I(y^{\alpha}(s),x^{i}(s))$, where $I(y^{\alpha}(s),x^{i}(s))$ is the coupling pair potential. 
As usually we assume that $I(s)$ is due to a short distance potential consequently $I(s)$ is proportional to the number $M_{S}$ of particles of $\mathcal T$ located near the surface of $S$ and to the number $N_{nn}$ of nearest neighbors of these particles but pertaining to $S$. Thus $I(s) \approx M_{S} N_{nn} \bar{I}(\alpha,i)$ where $\bar{I}(\alpha,i)$ is the mean value of the coupling pair potential. Since $M_{S} << M$ the properties of $\mathcal T$ can be calculated independently of $I(s)$ with a good approximation. \\
Since $\mathcal T$ is a large system at equilibrium the results obtained above show that we can associate to it a well defined temperature $T$, a characteristic time $\tau = \frac{\hbar}{k_{B} T} $, a length $\lambda^{2} = \frac{\hbar}{m_{\alpha}}\frac{\tau}{2}$ and its proper partition function 
\begin{equation}
Q_{\mathcal T} = \frac{1}{M!} \int dy^{M}(0) \int \mathcal{D} y^{M}(t) exp(- \frac{1}{\hbar} \int ^{\tau}_{0} H_{\mathcal T}(s)ds)
\label{T11}
\end{equation}
For $S$ coupled with $\mathcal T$ we introduce a new transition function independent on the origin of time $(t_{0} =0)$ but associated with a time interval $\tau$ in order to have the equilibrium for $\mathcal T$   
\begin{equation}
\frac{1}{Q_{\mathcal T}}\frac{1}{N!} \int\mathcal{D} x^{N}(t) exp(- \frac{1}{\hbar} \int ^{\tau}_{0} H(s) ds )\frac{1}{M!} \int dy^{M}(0) \int \mathcal{D} y^{M}(t) exp(- \frac{1}{\hbar} \int ^{\tau}_{0} (H_{\mathcal T}(s) + I(s))ds)
\label{T1}
\end{equation}
If all the potentials exhibit no significant variation on $\lambda$ we introduce 
\begin{equation}
exp - \beta \tilde{I}(0) = \frac{\int dy^{M}(0) exp(- \beta (U_{\mathcal T}(0) + I(0))}{\int dy^{M}(0) exp(- \beta U_{\mathcal T}(0))}
\label{coupling}
\end{equation}
and the integration of Eqs. (\ref{T1}) on $x^{N}(0)$ will produce the partition function of $S$ in contact with the thermostat, we have
\begin{equation}
Q_{S} = \frac{1}{N!}\int dx^{N}(0) \exp(-\beta (U_{S}(0) + \tilde{I}(0)))\int \mathcal{D} x^{N}(t) \exp(-\frac{1}{\hbar}\int^{\tau}_{0} K_{S}(s)ds)
\label{T2}
\end{equation}
in which $U_{S}$ and $K_{S}$ represent the total potential and kinetic energy for the system isolated. The term related to $K_{S}(s)$ corresponds to the ideal system, it can be calculated with Eqs. (\ref{fiideal}). Using the same value of $Cte$ as previously we get 
\begin{equation}
Q= \frac{1}{N!h^{3N}}\int dx^{N}(0) dp^{N}\exp{- \beta (U_{S}(0) + \tilde{I}(0) + \sum^{N}_{i=1} \frac{(p^{i})^{2}}{2m}})
\label{thermo}
\end{equation}
In our approach $\mathcal T$ defined the temperature and the time interval $\tau$ in addition the coupling between $S$ and $\mathcal T$ introduces an extra potential $\tilde{I}(0))$ defined in (\ref{coupling}). If the system $S$ is large in such a way that $N >> M_{S}N_{nn}$ we can neglect the contribution of $\tilde{I}(0))$ in comparison with the one of $U^{i}(0)$ and the traditional form of the partition function is re-obtained. \\
In the next Section we present the derivation of a macroscopic time-irreversible behavior. 

\section{Time-irreversible behaviors}
In Section $2C$ we have established that the transition function verifies a time-irreversible equation Eqs. (\ref{dfidt1}). On a simple case in Section $3.D$ we have introduced a probability verifying a Smoluchovski equation (\ref{smol}) which is the prototype of an equation describing a time-irreversible process. There is no doubt that with our approach we will be able to describe a large class of time-irreversible processes. In this Section we show how the presence of a bath may generate a friction phenomena on particles. We have already investigated such a kind of problem in [\cite{jpb5}], [\cite{jpb6}]. The interest of presenting a short summary of these works is double. First we show that a system in thermal equilibrium with a thermostat or in presence of a bath giving rise to a time-irreversible behavior can be treated on the same footing. Second we illustrate the difference between our approach and the so called system+reservoir approach. 

\subsection{A simple system}
In what follows the system $S$ will be reduced to one particle that we will call the Particle to be short, it obeys to the hamiltonian defined from Eqs.(\ref{K}) and Eqs.(\ref{U}) in which the interaction potential is reduced to an external potential. We assume that the Particle is embedded in a bath $B$ replacing the thermostat $\mathcal T$ investigated in the previous section. Now the interaction energy bath/Particle cannot be neglected in the Particle dynamics. In contrast the bath properties can be calculated in absence of the Particle. As in the previous Section we associate with the large bath a characteristic time $\tau$ a temperature $T$ and a distance $\lambda$. System and bath are enclosed in a volume $V$ and the initial position of the Particle will be used to define the center of the spatial coordinates. Many technical details concerning this model are given in [\cite{jpb5}].\\
Now we write the total hamiltonian as $H(s) = H_{S}(s) + H_{B}(s) + I(s)$ and a particular forms will be chosen for $B$ and $I$. 
The bath is represented by $M$ particles oscillating around their initial equilibrium positions, they behave as $M$ independent identical oscillators of frequency $\omega$ and of mass $m_{B}$. The hamiltonian associated with the bath is  
\begin{equation} 
H_{B}(t)= \sum^{M}_{i=1} [\frac{1}{2}m_{B} (\frac{dr_{i}(t)}{dt})^{2} +  \frac{1}{2} m \omega^{2} r_{i}^{2}(t)]
\label{hbath}
\end{equation} 
in which $r_{i}$ means for $i$ its deviation from its initial equilibrium position. For the interaction between bath and Particle we use the bi-linear hamiltonian frequently retained in the literature [\cite{jpb5}] 
\begin{equation}
I(t) = \sum^{M}_{i = 1} C_{i} r_{i}(t) x(t)
\label{hinter}
\end{equation}
in which $C_{i}$ is related to the second spatial derivative of the interaction potential between the Particle located at the center of the box at the initial time and the equilibrium position of the $i$ bath particle; $C_{i}$ depends on $i$ and for a short range Particle-bath interaction potential the sum in Eqs. (\ref{hinter}) is restricted to the first neighbors of the Particle. This potential requires the introduction of a renormalization constant that has been largely discussed in the literature [\cite{jpb5}] and it will be not considered here.  
The transition function that we have to consider is   
\begin{equation}
\phi(x_{0},r^{M}(0),t_{0}; x,r^{M}, t) = \frac{1}{M!}\smallint \mathcal{D}x(t) \exp{- \frac{1}{\hbar} \int\limits_{t_{0}}^{t} H_{S}(s)ds}
\smallint \mathcal{D}r^{M}(t) \exp{- \frac{1}{\hbar} \int\limits_{t_{0}}^{t}(H_{B}(s) + I(s))ds}
\label{newq1} 
\end{equation} 
In Eqs. (\ref{newq1}), $r^{M}(t)$ represents the set of the deviations from the initial positions 
as defined above, $r^{M}(t) = [r_{1}(t), ... r_{i}(t), ...r_{M}(t)]$.
Due to the quadratic form of $(H_{B}(s) + I(s))$ the functional integral on the 
variables $r^{M}(t)$ can be calculated exactly using a mathematical trick 
introduced in [\cite{feyn1}]. We write $r_{i}(t) = r_{i}(t)_{opt} + \delta r_{i}(t)$ 
in which $r_{i}(t)_{opt}$ corresponds to the optimization of the euclidean action associated 
with $(H_{B}(s) + I(s))$. Here the optimization is a mathematical procedure leading to 
a differential equation which is not the equation of motion, in contrast with what is 
done in pure quantum mechanics where the optimization of the lagrangian action is performed. We can write 
\begin{equation}
\smallint \mathcal{D}r^{M}(t) \exp{- \frac{1}{\hbar}\int\limits_{t_{0}}^{t}(H_{B}(s) + I(s))ds} = 
C_{opt} \exp{- \frac{1}{\hbar} [\int\limits_{t_{0}}^{t}(H_{B}(s) + I(s))ds}]_{opt} 
\label{newq2}
\end{equation}
in which the subscript $opt$ means that we have to calculate the trajectories on the 
optimum paths and $C_{opt}$ is a quantity independent of the particle positions, it 
results from the integration on the variable $\delta r_{i}(t)$. For two arbitrary 
times $t$ and $t_{1} > t$ for which the sets of positions 
correspond to $(x,r^{M})$ and $(x_{1},r^{M}_{1})$ respectively, we define 
\begin{equation}
\delta A[x,r^{M},t; x_{1},r^{M}_{1},t_{1}] = [\int\limits_{t}^{t_{1}} ((H_{B}(s) + I(s))ds]_{opt}
\label{deltaA}
\end{equation}
Using the hamiltonians Eqs. (\ref{hbath}) and Eqs. (\ref{hinter}) we can get the explicit 
form of $\delta A[x,r^{M},t; x_{1},r^{M}_{1},t_{1}]$ giving
\begin{equation}
\phi(x_{0},r^{M}(0),t_{0}; x,r^{M}, t) = C_{\delta} \smallint \mathcal{D}x(t) \exp{- \frac{1}{\hbar}[A[x_{0},t_{0}; x,t] + \delta A[x_{0},r^{M}(0),t_{0}; x,r^{M},t]]}
\label{newq4}
\end{equation}
and the transition function for the small system $\bar{\phi}(x_{0},t_{0}; x, t)$ can be written
\begin{equation} 
\bar{\phi}(x_{0},t_{0}; x, t)= C_{\delta} \smallint \mathcal{D}x(t) \exp{- \frac{1}{\hbar}A_{P}[x_{0},t_{0}; x,t]}
<\exp{- \frac{1}{\hbar}\delta A[x_{0},r^{M}(0),t_{0}; x,r^{M},t]}>_{bath} 
\label{qbar}
\end{equation}
in which $C_{\delta}$ is a normalization constant and $< ... >_{bath}$ means that we have to take a procedure in order to eliminate 
the bath particle positions. This procedure depends on the physics under consideration.\\ 
We assume that the bath is in thermal equilibrium in the field created by the Particle for any value of $t$. 
To describe the bath equilibrium we focus on closed paths explored during a time interval $\tau$. The time interval $(t - t_{0})$ is sliced in equal intervals of thickness $\tau$ and we assume that $\tau$ is vanishingly small in comparison with $(t - t_{0})$, this corresponds to a well common assumption, the Smoluchowski or Fokker- Planck equations do not describe short ranged processes. Finally we assume that $\omega \tau << 1$ a physical reasonable approximation taking into account that $\tau \approx 10^{-14}sec$ at room temperature. With such conditions we may show that 
$<\exp{- \frac{1}{\hbar}\delta A[x_{0},r^{M}(0),t_{0}; x,r^{M},t]}>_{bath}$ contains two parts one produced a renormalization of the interaction potential and the second one is given by 
\begin{equation}
\frac{-1}{\hbar} \frac{\sum^{M}_{i=1}C_{i}^{2}}{4 m_{B} \omega^{2}} \tau^{2}\int^{t}_{t_{0}} (\frac{dx}{ds})^{2}ds
\label{correct}
\end{equation}
This term has to be added to the kinetic part of $H_{S}$ this is equivalent to replace the mass of the Particle by an effective mass given by
\begin{equation}
m_{eff}= m + \frac{\sum^{M}_{i=1}C_{i}^{2} \tau^{2}}{2m_{B} \omega^{2}}
\label{effmass}
\end{equation} 
If the interaction is short ranged we have $\sum^{M}_{i=1}C_{i}^{2} \approx N_{nn} C$ in which $N_{nn}$ is the mean number of nearest neighbors at the initial time and $C$ the common value of the coupling constant. We introduce a characteristic length for the vibration $l$ to which we associate a mean value for the vibrational energy $E_{vib} = \frac{1}{2}m_{B}\omega^{2} l^{2}$ and we write the interaction energy as $E_{int} = C l \lambda$. The effective mass (\ref{effmass}) can be rewritten as 
\begin{equation}
m_{eff} = m [1 + N_{nn} \frac{m_{B}}{m} \frac{E_{int}^{2}}{E_{vib} E_{kin}}]
\label{masseff}
\end{equation} 
in which $E_{kin} = \frac{1}{2} k_{B}T$. We see that $m_{eff} \to m$ if the ratio $N_{nn} \frac{m_{B}}{m} \frac{E_{int}^{2}}{E_{vib} E_{kin}}$ is very small, this can be realized if the bath particles are extremely light $(\frac{m_{B}}{m} << 1)$ or if the coupling energy is extremely weak in comparison of the proper energies of the bath particles $E_{vib}$ and $E_{kin}$. In opposite $m_{eff}$ is very different from $m$ if the bath particles are very heavy in comparison with $m$ or in the case of a very strong coupling. All these results are expected from our general knowledge concerning the brownian motion. More generally from Eqs. (\ref{masseff}) we may observe a smooth transition from the original quantum system to a classical one while in all cases a Smoluchovski equation will be associated with the system evolution [\cite{jpb5}]. We consider the existence of the Smoluchovski equation as the signature of an irreversible behavior. \\

\subsection{More sophisticated models and other approaches}
In the previous subsection we have used our formalism to predict one kind of irreversible behavior that does not require the introduction of any new basic assumptions in comparison with those needed for describing a thermal equilibrium state. Analytical results have been obtained resulting from simple assumptions. Some extensions can be easily introduced, for instance, $S$ can be formed by a given number of interacting particles instead of only one, the different time scales can be modified or a more sophisticated bath particle interaction can be considered. In a recent paper [\cite{jpb6}] we have investigated an example in which the Particle is not in equilibrium with its surrounding then a memory effect is introduced in the transition function and the dynamics of the Particle cannot be separated from the one of bath particles. Simple approximations show that we keep the irreversible behavior of the system as expected. \\
There is a large literature devoted to the  coupling between a small system and a bath in view to derive a Langevin or a Fokker-Planck type equation for brownian particles (see for instance [\cite{caldeira}],[\cite{grabert}], [\cite{weiss}], [\cite{datta}]). In these 
approaches the ensemble system+reservoir is at the equilibrium and described from a density matrix, the small system considered as an isolated system obeys to a Schr\"{o}dinger equation and the irreversibility arises from the energy transfer from the small system to its large environment. Here the transition function Eqs. (\ref{tool}) gives a quantum description of a N-body system and a time-irreversibility behavior exists even for small quantum systems in absence of bath. This last point is of great interest in order to describe for instance the time evolution of an irreversible chemical reaction. Moreover we are able to describe the smooth transition from a time-irreversible quantum regime to a time-irreversible classical regime. The mathematical apparatus used here is different and simpler that the one used in system+reservoir approaches. \\
It is not obvious that the formalism developed here will be also efficient to describe a time-irreversibility having another origin but at least it is sufficient for describing the time-irreversibility of an important class of models.

\section{Relation with a pure quantum mechanical regime}
In the previous Sections we studied the connection between statistical mechanics and thermodynamics but besides thermodynamics it exists a hamiltonian mechanics in which there is no friction phenomena and all the observed process are time-reversible. Hereafter we extend our approach to describe such reversible processes. This leads to investigate the connection between $ST$ and the quantum mechanics described by the Schr\"{o}dinger equation $(SE)$.

\subsection{Reversible mechanics}
Classical mechanics describes reversible process because the Newton law is invariant if we change $t$ into $-t$ [\cite{arnold}]. For a given hamiltonian a particle starts from $x_{0}$ at the time $t_{0}$ and reaches $x_{1}$ at the time $t_{1}$ following a well defined trajectory. Reversibility means that starting from $x_{1}$ we can describe the same trajectory if $t$ is changed in $-t$. In quantum mechanics the situation is no so simple. The Schr\"{o}dinger equation [\cite{cohen}]
\begin{equation}
i\hbar \frac{\partial \psi(x,t)}{\partial t} = - \frac{\hbar^{2}}{2m} \frac{\partial ^{2} \psi(x,t)}{\partial x^{2}} + V(x,t) \psi(x,t)
\label{es}
\end{equation}
describes the time evolution of the wave function $\psi(x,t)$ for a system in presence of the external potential $V(x,t)$. This equation contains a first derivative relative to $t$ suggesting that it might describe time-irreversible behaviors. However this is not correct because only the product $\psi(x,t).\psi^{*}(x,t)$ in which $\psi^{*}(x,t)$ is the complex conjugate of $\psi(x,t)$ has a physical meaning. In addition to Eqs. (\ref{es}) we have to consider a second equation
\begin{equation}
-i\hbar \frac{\partial \psi^{*}(x,t)}{\partial t} = - \frac{\hbar^{2}}{2m} \frac{\partial ^{2} \psi^{*}(x,t)}{\partial x^{2}} + V(x,t) \psi^{*}(x,t)
\label{es*}
\end{equation} 
obtained by taking the complex conjugate of Eqs. (\ref{es}); in comparison with Eqs. (\ref{es}) no new physics is introduced. If we change $\psi(x_{1}, t_{1})$ into $\psi^{*}(x_{1}, t_{1})$ and consider the backward evolution we obtain $\psi^{*}(x_{0}, t_{0})$ which is nothing else than the complex conjugate of $\psi(x_{0}, t_{0})$. The $SE$ is said time-reversible in the Wigner sense (see[\cite{maes}]). In order to prove the conservation of the probability $\psi(x,t).\psi^{*}(x,t)$ we have to use simultaneously Eqs. (\ref{es}) and Eqs. (\ref{es*}) [\cite{cohen}]. 
 
\subsection{The backward transition function}
In the previous Sections $\phi(x^{N}(0), t_{0}; x^{N}, t)$ was defined for $t \geq t_{0}$ and from the thermodynamic point of view the reverse motion is impossible. If we forget thermodynamics and focus on pure reversible mechanics we can force the system to be time-reversible. This will be done in the spirit developed just above. In the time interval $(t_{0},t_{1})$ in addition to the forward transition function $\phi(x^{N}(0), t_{0}; x^{N}, t)$ verifying Eqs. (\ref{dfidt1}) we introduce a backward transition function $\hat{\phi}(x^{N},t;x^{N}(1),t_{1})$ defined for $t \leq t_{1}$. As in the previous subsection we want to describe the reverse motion with the same ingredients that we have used for the forward motion. As a first step we assume the existence of a semi-group property defined for the forward motion. Similarly to Eqs. (\ref{semigroup}) we write
\begin{equation}
\hat{\phi}(x^{N}, t-\epsilon;x^{N}(1), t_{1}) = \int dx'^{N}\hat{\phi}(x^{N}, t-\epsilon; x'^{N},t ) \hat{\phi}(x'^{N},t; x^{N}(1), t_{1})
\label{semieps2}
\end{equation}
This relation is well defined since we have $(t- \epsilon \leq t_{1})$, and $(t -\epsilon \leq t)$. In a second step we establish a connection between forward and backward motions. In quantum mechanics the evolution operator of $\psi(x,t)$ for a time interval (t- t') is identical to the complex conjugate operator acting on $\psi^{*}(x,t)$ provided we consider the inverse time $(t'-t)$. We can translate this idea in our case. For the infinitesimal time interval $\epsilon$ we assume that $\hat{\phi}(x^{N}, t-\epsilon; x'^{N},t )$ calculated for the time running backward from $t$ to $t - \epsilon$ is equal to ${\phi}(x^{N}, t-\epsilon; x'^{N},t )$ for which the time is running forward from $t-\epsilon$ to $t$ and Eqs. (\ref{semieps2}) will be written
\begin{equation}
\hat{\phi}(x^{N}, t-\epsilon;x^{N}(1), t_{1}) = \int dx'^{N}{\phi}(x^{N}, t-\epsilon; x'^{N},t ) \hat{\phi}(x'^{N},t; x^{N}(1), t_{1})
\label{semieps}
\end{equation}
Arguments identical to those used to derive Eqs. (\ref{dfidt1}) lead to the following differential equation
\begin{equation}
- \frac{\partial\hat{\phi}(x^{N}, t; x^{N}(1),t_{1})}{\partial t} = \frac{\hbar}{2m} \sum^{N}_{i=1} \frac{\partial^{2} \phi(x^{N}, t; x^{N}(1), t_{1})}{\partial (x^{i})^{2}} 
- \frac{1}{\hbar} \sum^{N}_{i=1} U^{i}(t)\phi(x^{N}, t; x^{N}(1), t_{1}) 
\label{hatdfidt}
\end{equation} 
The function $\hat{\phi}(x^{N},t)$ defined according to
\begin{equation}
\hat{\phi}(x^{N},t) =\int \hat{\phi}(x^{N},t;x^{N}_{1},t_{1})\hat{\phi}_{1}(x^{N}_{1})dx_{1}
\label{deffihat}
\end{equation}
exhibits a time dependence that we can be obtained in the same way as for $\phi(x,t)$ and we get 
\begin{equation}
-\frac{\partial\hat{\phi}(x^{N},t)}{\partial t} = \frac{\hbar}{2m} \sum^{N}_{i=1} \frac{\partial^{2} \hat{\phi}(x^{N},t)}{\partial (x^{i})^{2}} 
- \frac{1}{\hbar} \sum^{N}_{i=1} U^{i}(t)\hat{\phi}(x^{N},t) 
\label{dfidt3}
\end{equation}

\subsection{Relation with the Schr\"{o}dinger equation}
To relate the previous results with the $SE$ we take $N=1$.
From the pair of non-negative functions $\phi(x,t)$ and $\hat \phi(x,t)$ we associate two new functions $R(x,t)$ and $S(x,t)$ defined by  
\begin{equation}
R(x,t)= \frac{1}{2} \ln {\phi(x,t)\hat \phi(x,t)}
\label{Rxt}
\end{equation}
and
\begin{equation} 
S(x,t) = \frac{1}{2} \ln{\frac{\hat{\phi}(x,t)}{\phi(x,t)}}
\label{Sxt}
\end{equation}
From $R(t,x)$ and $S(t,x)$ we may define a complex valued functions and its complex conjugate 
\begin{equation}
\psi(x,t) = \exp{[R(x,t) - i S(x,t)]}
\label{psi}
\end{equation}
\begin{equation}  
\psi*(x,t) = \exp{[R(x,t) + i S(x,t)]}
\label{psi*}
\end{equation}
It is clear that we have $ \psi(x,t).\hat{\psi}^{*}(x,t) =\phi(x,t).\hat{\phi}(x,t)$. Straightforward mathematical calculations [\cite{naga1}] show that $\psi(x,t)$ verifies the  Schr\"{o}dinger equation (\ref{es}) the potential $V(x,t)$ being related to $U(x,t)$ by the relation
\begin{equation}
V(x,t) - U(x,t) + 2 \hbar [{\partial S(t,x)}/{\partial t} + \frac{\hbar}{2m}(\nabla_{x}S(t,x))^{2}] = 0
\label{vetu}
\end{equation} 
Thus, by adding to the transition function Eqs. (\ref{tool}) describing the forward motion a transition function describing the reverse motion we may define a complex valued function verifying a Schr\"{o}dinger like equation provided the potentials introduced in the various equations are related according to Eqs. (\ref{vetu}). \\
These results have to be understand as follows. Transformations like Eqs. (\ref{rets}) have been used in the literature, for instance, by [\cite{nelson}] and [\cite{bohm}] essentially in order to modify the $SE$ (see for instance [\cite{omnes}] for a discussion) but this is not our goal we consider that 
the couple $(\phi(x,t),\hat{\phi}(x,t))$ with the equations Eqs. (\ref{dfidt2}), (\ref{dfidt3}) and (\ref{vetu}) produce a quantum description mathematically equivalent to the traditional one given by $(\psi(x,t), \psi^{*}(x,t))$ 
with the equations (\ref{es}) and (\ref{es*}). For each description we have to solve a linear differential equation from which we obtain $R(t,x)$ and $S(t,x)$ then we also get the result in the second description. If the first description is the $SE$ one, for instance, we we use $V(x,t)$ to get $R(t,x)$ and $S(t,x)$ from them and (\ref{vetu}) we obtain $U(x,t)$ as a function of $x$ and $t$ only. We can verify that $R(t,x)$ and $S(t,x)$ give rise to the description in terms of $(\phi(x,t),\hat{\phi}(x,t))$ with the potential $U(x,t)$. In (\cite{badia}) we will show the interest to have two equivalent descriptions and that the relation (\ref{vetu}) is meaningful.  \\

\section{Conclusions}
In this work our main goal was not to challenge the usual methods of calculation in statistical mechanics but to show that the problem of time-irreversibility in $ST$ can be formulated with a new point of view. Thermodynamics shows that all real processes transforming an equilibrium state into another one are time-irreversible in reality. From this fact we decide to found $ST$ in such a way that equilibrium states and time-irreversible processes can be described on the same footing. This leads to a first important point: the Schr\"{o}dinger that only describes reversible processes can not be the corner stone on which statistical thermodynamics can be found. Here the quantum physics is described via a transition function that is a time-dependent transformation of the hamiltonian via the path integral formalism. We show that this function verifies a time-irreversible partial differential equation to which we can associate in general condition a Smoluchovski type equation. This is the second important point of our work. The traditional problematic in which we try to conciliate a reversible behavior at a microscopic level with a time-irreversible behavior at a macroscopic level is inverted since we start from a quantity which is time-irreversible at a microscopic level. Then it is crucial to the prove that our dynamical approach may lead to stable equilibrium states and also that we can recover the traditional results obtained via the Gibbs ensemble method. From the transition function we have to our disposal a path probability from which we can analyze several properties concerning the closed paths on which we focus: mean volume, kinetic energy and time needed to explore the paths. We show that the quantum physics introduces a unit of time characterizing an equilibrium state. From the analysis of the energy fluctuations on the paths a link has been established with thermodynamics and all the traditional results have been obtained for very large systems or for system in contact with a thermostat. If the thermostat is replaced by a bath without no new approximation we describe a time irreversible behavior of the system via the existence of a Smoluchovski equation. Our approach is different from the so called system+reservoir approach since the time-irreversibility is not associate with the presence of the bath and we can predict a smooth transition from quantum to classical domain. 
Of course the use of a transition function is not restricted to the description of thermodynamic processes. This has been illustrated on two different aspects. First it has been already shown that we can investigate the behavior of small systems for which there is no thermodynamics, an illustrative example corresponds to the description of an irreversible chemical reaction in vacuum. Secondly we can generate a time-reversible behavior in order to describe systems obeying to hamiltonian mechanics. Our approach has been implemented by the existence of a second transition function describing the reverse motion. Instead of dealing with two real non-negative transition functions we can combine them into a complex valued function verifying a Schr\"{o}dinger equation. This requires to have a potential associated with each representation. This corresponds to the third important result obtained in this work.\\
It was possible to start our work from the Schr\"{o}dinger equation, to split it in two transition functions and to decide to keep only the one related to the forward motion in order to break the time-reversibility of this equation. This point shows that the difference between statistical thermodynamics and quantum mechanics must be analyzed in term of time reversibility/irreversibility. It explains why the wave function is necessarily a complex valued function while in statistical physics it is sufficient to consider a real valued function having a physical meaning by itself. In this work we decided to ignore the Schr\"{o}dinger equation from the very beginning preferring to follow a route that seems more constructive and more direct. This choice also results from Feynman remarks (\cite{feyn1}) suggesting that it probably exists a way to derive the path integral approach in statistical physics directly from the path integral description for the time dependent motion without using the total apparatus associated with the Schr\"{o}dinger equation.


\end{document}